\documentclass[9pt,twocolumn,twoside]{opticajnl}
\journal{opticajournal} 
\DeclareMathOperator{\re}{Re}

\newcommand{\oma}{\omega_\mathrm{A}}
\newcommand{\Dav}{\overline{\Delta}}
\setboolean{shortarticle}{true}

\usepackage[numbers]{natbib}
\usepackage{upgreek}

\usepackage{pdfpages}

\usepackage{lineno}

\title{Precision of the acoustic control of single photon scattering with semiconductor quantum dots}

\author[ ]{Rafał A. Bogaczewicz}
\author[*]{Paweł Machnikowski}

\affil[ ]{Institute of Theoretical Physics, Wrocław University of Science and Technology, 50-370 Wrocław, Poland}

\affil[*]{pawel.machnikowski@pwr.edu.pl}

\begin{abstract}
Acoustic modulation of quantum dots allows one to control the scattering of photons. Here we theoretically characterize the degree of this acoustic control in the frequency domain. We formulate the theory of low-intensity resonance fluorescence in the presence of white noise and show that a high level of control is achievable with a two-tone acoustic field for appropriate settings of modulation amplitudes as long as the noise-induced phase diffusion coefficient remains one order of magnitude smaller than the acoustic frequency. In addition, using a quantitative model of optical signal collection, we determine that the acoustic phase must be stable over $\mathbf{10^4}$ to $\mathbf{10^5}$ acoustic periods for efficient control. 
\end{abstract}

\setboolean{displaycopyright}{false} 

\begin{document}

\maketitle

\section{Introduction}
\label{sec:introduction}

Resonance fluorescence (RF), which gives rise to non-classical light with special coherence properties \cite{Dalibard1983,Nienhuis1993}, has recently found applications in solid-state systems \cite{Delley2017,muller_resonance_2007,wrigge_efficient_2008,Scholl2019}.
Modulation of a quantum emitter offers additional control of the light scattering process \cite{Trivedi2020,Ilin2023,DeCrescent2024}. In the resolved sideband regime \cite{Metcalfe2010,Villa2017}, the RF spectrum of a weakly optically excited and acoustically modulated semiconductor quantum dot (QD) contains the usual central line, located at the laser frequency \cite{ScullyZubairy1997}, and a series of sidebands induced by modulation. The scattered photons are antibunched \cite{Villa2017,Weiss2021}, indicating the single-photon nature of the scattering.

Surface acoustic wave (SAW) mixing has been shown to provide precise control over the scattering of photons to a particular frequency channel \cite{Weiss2021}, as well as in the time domain \cite{Wigger2021}. 
Two-tone acoustic driving leads to quantum interference of different pathways to a given scattering process that involve the two acoustic harmonics in various combinations. This interference is governed by the relative phase between the two acoustic waves, which allows one to control the probability of photon scattering to a given frequency sideband \cite{Weiss2021}. 
Theoretical description has been extended to quantum acoustic modes \cite{Hahn2022,Groll2023}, which opens up a perspective for the implementation of frequency- and time-bin encoding \cite{Pan2012,Lu2023}, quantum multiplexing \cite{Piparo2019}, or quantum acousto-optic transduction \cite{Stannigel2010}. 
Short wavelengths of acoustic waves in the GHz  frequency range make them perfect candidates for miniaturized devices that may lay the ground for on-chip acousto-optic quantum hybrid systems \cite{Weiss2018,Delsing2019}.

Whether this acoustic control can be exploited in quantum applications depends, to a large extent, on the resilience of the observed coherent acoustooptic features against external noise that leads to random fluctuations of the transition energy. In addition, in view of the finite time required to collect the optical signal originating from a single quantum emitter, phase stability of the control fields becomes crucial.

In this paper, we theoretically analyze the achievable degree of control (DOC) of photon scattering by coherent acoustic modulation of a QD. We develop a model of low-excitation RF of a periodically modulated two-level system in the presence of white noise. This allows us to determine the scattering spectrum and to show that the relative contrast of phase-dependent intensity oscillations at the optimal setting of modulation parameters is weakly affected by noise as long as the strength of the latter remains well below the acoustic frequency. Finally, we set the minimum requirements for the stability of the acoustic frequency in these coherent acoustooptic processes. 

\section{Model}
\label{sec:model}

We consider a self-assembled semiconductor QD resonantly driven by a weak, monochromatic laser field. The scattered photons are collected and their time-integrated spectrum is determined \cite{Weiss2021}. The QD transition energy is modulated by a SAW via deformation-potential coupling to crystal strain and is subject to random environmental noise. Our general model can be applied, e.g., to a typical InGaAs QD with several meV separation between the fundamental transition addressed here and excited levels and the exciton life time of 1~ns. The doublet of optically active transitions can be resolved by light polarization, leading to an effectively two-level system. Typical acoustic frequencies used in experiments are in the range of hundreds of MHz. In view of the relatively slow  acoustic modulation and exciton decay, we assume Gaussian noise with negligible correlation time, i.e. white noise. This model corresponds to the short-memory limit of a bath of oscillators with Ohmic spectral density, as well as to a sum of a large number of fast telegraph noise sources, like charge traps typical for solid-state environments \cite{Freeman2016,Kuhlmann2013} (see Supplement 1). An approximation of such a process could also, in principle, be generated artificially using radio-frequency electronics. As a rough measure of typical noise strengths one could take the fluctuation-induced single-QD line widths, which are on the order of 10~GHz \cite{Kuroda2010}.

We model the QD as a two-level system with energy eigenstates $|0\rangle$ and $|1\rangle$ and transition energy $\hbar\omega_{0}(t)$, where the time dependence results from SAW modulation and  noise,
\begin{equation}
\omega_0(t) = \overline{\omega}_0 + \Delta\omega_\mathrm{ac}(t) + \Delta\omega_\mathrm{ns}(t), \label{eq:transition_frequency} 
\end{equation}
where  $\overline{\omega}_0$ is the unperturbed transition energy, whereas $\Delta\omega_\mathrm{ac}(t)$ and $\Delta\omega_\mathrm{ns}(t)$ denote acoustic modulation and noise contributions, respectively, both with zero average. We assume that $\Delta\omega_\mathrm{ac}(t)$ is periodic with fundamental frequency $\oma$. The system undergoes spontaneous emission with the rate $\gamma$.

The evolution is found by iteratively solving the Lindblad equation for the density matrix up to the second order in Rabi frequency, following Refs.~\cite{Weiss2021,Wigger2021}. 
The detector response is described by the autocorrelation function
$G(t_1,t_2) = \left\langle \sigma_+(t_1)\sigma_-(t_2) \right\rangle$,
which is calculated using the quantum regression theorem \cite{Bogaczewicz2023, Wigger2021}. In the leading order in the field amplitude (~$\Omega^2$) it reads for $t_2>t_1$ \cite{Weiss2021}
\begin{align}
G(t_1,t_2) = & \frac{\Omega^2}{4}
\int_{-\infty}^{\infty} du 
e^{-\left(\frac{\gamma}{2}+i\overline{\Delta}\right)u} 
e^{-i\Phi_\mathrm{ac}(t_1,t_1-u)}
\label{eq:autocorrelation_2} \\
& \times \int_{-\infty}^{\infty} du' 
e^{-\left(\frac{\gamma}{2}-i\overline{\Delta}\right)u'}
e^{i\Phi_\mathrm{ac}(t_2,t_2-u')} 
\mathcal{D}(u,u',t_2-t_1), \nonumber
\end{align}
where $\Phi_\mathrm{ac}(t_2,t_1)=\int_{t_1}^{t_2}\Delta\omega_\mathrm{ac}(s)ds$ is the deterministic contribution to the accumulated phase, $\overline{\Delta}=\omega_\mathrm{L}-\overline{\omega}_0$, with $\omega_\mathrm{L}$ denoting the laser frequency, and 
\begin{equation}\label{eq:D}
\mathcal{D}(u,u',\tau)=\overline{e^{i\Phi_\mathrm{ns}(0,-u)-i\Phi_\mathrm{ns}(\tau,\tau-u')}}\theta(u)\theta(u')  
\end{equation}
encodes the complete information about the noise, which is assumed stationary, with 
$\Phi_\mathrm{ns}(t_2,t_1)=\int_{t_1}^{t_2}\Delta\omega_\mathrm{ns}(s)ds$,
$\theta(u)$ denoting the Heaviside step function and the bar representing averaging over noise realizations. Note that $\Phi_\mathrm{ac}(t_2,t_1)$ is periodic in both arguments.

White noise leads to phase diffusion described by a Gaussian distribution for $\Phi_\mathrm{ns}(t_2,t_1)$ with variance $2D(t_2-t_1)$, $t_2>t_1$, where $D$ is the diffusion coefficient related to the noise strength, $\langle \omega_\mathrm{ns}(t)\omega_\mathrm{ns}(t+\tau)\rangle = 2D\delta(\tau)$. For $\tau>0$ we obtain (see Supplement 1) $ 
D(u,u',\tau) = \mathcal{D}_\mathrm{el}(u,u',\tau) + \mathcal{D}_\mathrm{inel}(u,u',\tau)$,
where 
\begin{equation}
\mathcal{D}_\mathrm{el}(u,u',\tau) = e^{-D(u+u')}\theta(u)\theta(u') \label{eq:D_el}
\end{equation}
and 
\begin{align}
\mathcal{D}_\mathrm{inel}(u,u',\tau) = & e^{-D\tau} \theta(u)\theta(u'-\tau) \label{eq:D_inel} \\
& \times \left[e^{-D|u+\tau-u'|}-e^{-D(u+u'-\tau)}\right]. \nonumber
\end{align}
$\mathcal{D}_\mathrm{el}$ does not depend on $\tau$, resulting in a periodic autocorrelation function that leads to a series of unbroadened spectral features. We refer to this contribution as elastic because the narrow (laser-limited) lines mean no energy exchange with the environment. $\mathcal{D}_\mathrm{inel}$ is damped in $\tau$, leading to broadened spectral lines, corresponding to inelastic scattering, in which the scattered photon looses or gains some energy. This contribution vanishes as $D\to 0$, hence it is fully due to noise.

Substituting \eqref{eq:D_el} and \eqref{eq:D_inel} to Eq.~(\ref{eq:autocorrelation_2}) one gets the corresponding elastic and inelastic contributions to the autocorrelation function. A function periodic in its two arguments is also periodic in their sum and difference. We can therefore define functions $\phi_n(u)$ through the expansion
\begin{equation}
e^{i\Phi_\mathrm{ac}(t-u',t-u)} = \sum_n \phi_n(u-u')e^{in\oma \left(t-\frac{u+u'}{2} \right)}.\label{eq:exp_Phi_1}
\end{equation}
For the elastic term, we then immediately find
\begin{equation}
G_\mathrm{el}(t_1,t_2) = \frac{\Omega^2}{\gamma^2}
\sum_{nm} c_nc_m^* e^{i\oma(nt_1-mt_2)},\quad t_2>t_1, \label{eq:autocorrelation_el}
\end{equation}
where
\begin{equation}\label{eq:c_n}
c_n = \frac{\gamma}{2} 
\int_0^{\infty}du e^{-(\gamma/2+D+i\Dav)u} e^{-in\oma u/2} \phi_n(u).
\end{equation}
The inelastic part can be written in the form (see Supplement 1)
\begin{align}  \label{eq:autocorrelation_inel}
G_\mathrm{inel}(t_1,t_2) & = \frac{\Omega^2}{2\gamma^2}
e^{-(\gamma/2+D-i\Dav)(t_2-t_1)} \\
& \times \sum_{nmk} b_m^* b_{n-k} d_k e^{i\oma(nt_1-mt_2)},\quad t_2>t_1, \nonumber
\end{align}
where $b_n$ are the coefficients of the Fourier expansion
\begin{equation}
e^{-i\Phi_\mathrm{ac}(t_2,t_1)} = \sum_{nm} 
b_m^{*} b_n e^{i\oma(nt_1-mt_2)} \label{eq:exp_Phi_2}
\end{equation}
and
\begin{equation}
d_n = \frac{2D\gamma}{(\gamma+D+in\oma)^2-D^2} 
\left(c_n + c_{-n}^* \right). \label{eq:d_n} 
\end{equation}

Both components of the autocorrelation function, as functions of $\tau$, are sums of purely exponential contributions. Consequently, the time-integrated RF spectrum,
\begin{equation}    \label{eq:RF_spectrum}
S(\omega) = \re  \int_0^{\infty}d\tau e^{i(\omega-\omega_L)\tau}e^{-\Gamma \tau} \overline{G}(\tau),
\end{equation}
where 
\begin{equation}
\overline{G}(\tau) = \frac{\oma}{2\pi} \int_0^{2\pi/\oma} dt G(t, t+\tau), \label{eq:autocorrelation_time_integrated}
\end{equation}
has an elastic and inelastic part, both composed of a series of Lorentzians (explicit formulas are given in Supplement 1). The former is only broadened by the finite instrumental resolution $\Gamma$ which we have included in the model. The spectral features of the latter are not only broadened by $\gamma/2+D$ but also shifted by $\Dav$, that is, from the spectral position related to the laser frequency to that bound to the unperturbed transition energy.

\section{Results}
\label{sec:results}

We set $\gamma/\oma = 2\Gamma/\oma = 0.1$. RF spectra will be presented in natural units set by the maximum value of the spectrum for an unperturbed system without modulation under weak resonant excitation $S_0 = \Omega^2/(\gamma^2\Gamma)$ \cite{ScullyZubairy1997}. Similarly, the natural unit for the intensity is $I_0=\pi\Omega^2/\gamma^2$, corresponding to the standard RF of an unperturbed weakly excited two-level system.

\subsection{RF spectrum under harmonic modulation and noise}

\begin{figure}[tb]
\includegraphics[width=\linewidth]{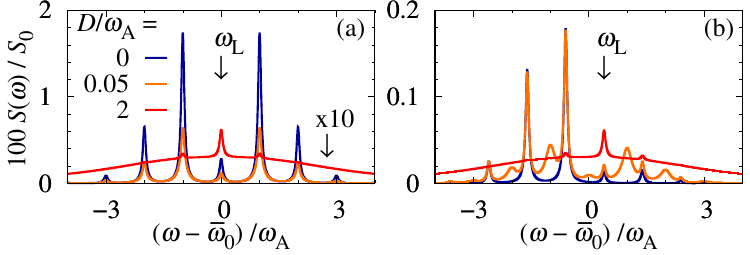}
\caption{\label{fig:RF_spectrum_single_SAW} RF spectrum for acoustically modulated QD with noise. Results for a single-SAW mode in case of (a) resonant excitation and for (b) $\omega_\mathrm{L} = \overline{\omega}_0+0.4\oma$.}
\end{figure}

Substituting Eq.~(\ref{eq:autocorrelation_el}) and Eq.~(\ref{eq:autocorrelation_inel}) to Eq.~(\ref{eq:autocorrelation_time_integrated}) and Eq.~(\ref{eq:RF_spectrum}) we find the spectrum as a sum of Lorentzian contributions as well as dispersive terms in the inelastic component.
We start with the case of harmonic modulation 
$\Delta\omega_\mathrm{ac}(t)=A\oma\cos(\oma t)$. Then
\begin{equation}
\label{eq:b_phi_1}
b_n^{(1)} = J_n(A),\quad 
\phi_n^{(1)}(u) = i^nJ_n\left[2A\sin\left(\oma u/2\right)\right],
\end{equation}
where $J_n$ is the Bessel function of the first kind. From this, $c_n$ and $d_n$ are calculated numerically using \eqref{eq:c_n} and \eqref{eq:d_n}.

Fig.~\ref{fig:RF_spectrum_single_SAW} presents the RF spectra in this case for resonant (Fig.~\ref{fig:RF_spectrum_single_SAW}a) and slightly detuned (Fig.~\ref{fig:RF_spectrum_single_SAW}b) excitation for $A=2$. In the absence of noise (blue lines), the spectrum consists of a series of lines separated from the laser frequency by an integer multiple of $\oma$ \cite{Weiss2021,Wigger2021}. 
Upon including noise (orange and red lines), the intensities of these peaks change. In addition, an inelastic contribution appears at integer multiples of $\oma$ from the unperturbed \textit{transition} frequency, similar to the case without modulation \cite{Bogaczewicz2023}. This is particularly well visible in Fig.~\ref{fig:RF_spectrum_single_SAW}b, where the laser is detuned by a fraction of acoustic frequency and the two line series appear at different frequencies. A similar inelastic contribution appears in the quantum regime, where the spectral jitter is modeled as pure dephasing \cite{Groll2023}. While the elastic lines remain narrow, the width of the inelastic ones grows, as discussed in Sec.~\ref{sec:model} until, for a sufficiently strong noise, they merge into a broad feature that dominates the spectrum (red lines). 

\begin{figure}[tb]
\includegraphics[width=\linewidth]{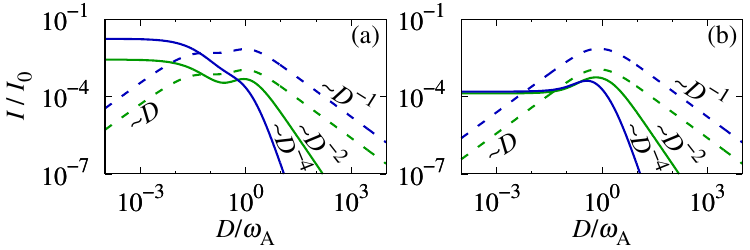}
\caption{\label{fig:Intensity_single_SAW} Intensities of lines from elastic/inelastic series (solid/dashed lines) for $p=q=0$ (green lines) and $p=q=1$ (blue lines). (a) Resonant excitation; (b) $\omega_\mathrm{L} = \overline{\omega}_0 + 0.4\oma$.}
\end{figure}

The intensity of an elastic line located at $\omega=\omega_\mathrm{L}+p\oma$ is
\begin{equation}
I_\mathrm{el}^{(p)} = I_0 |c_p|^2, \label{eq:I_el_p}
\end{equation}
while for an inelastic contribution at $\omega=\overline{\omega}_0+q\oma$ it is
\begin{equation}
I_\mathrm{inel}^{(q)} = I_0 \re b_q^*\sum_k b_{q-k} d_k /2. \label{eq:I_inel_q}
\end{equation}
Note that the latter can be interpreted as a line intensity only when $D\ll \oma$; otherwise the lines lose their identity. 
The intensities of selected contributions are shown as functions of the phase diffusion coefficient $D$
in Fig.~\ref{fig:Intensity_single_SAW}, where the green lines correspond to the central line at resonant excitation, while blue lines represent the intensities of the first sideband of the laser frequency and transition frequency for the elastic and inelastic contributions, respectively. Figs.~\ref{fig:Intensity_single_SAW}a,b correspond to the excitation conditions of Figs.~\ref{fig:RF_spectrum_single_SAW}a,b, respectively.
One can see that the intensities have a power-law asymptotic dependence on $D$. In the limit of vanishing noise, the intensities of the elastic lines (solid lines in Fig.~\ref{fig:Intensity_single_SAW}) reach a finite value, corresponding to the $D\to 0$ limit of $c_n$. In this limit, the intensities of the inelastic lines vanish proportionally to $D$, as follows from Eq.~(\ref{eq:d_n}), restoring the purely elastic scattering of the noise-free regime \cite{Weiss2021}. For strong noise ($D\gg \gamma,\oma,\overline{\Delta}$), all the intensities decrease with $D$. From Eq.~(\ref{eq:c_n}) one finds $c_n\sim D^{-(n+1)}$ (see Supplement 1), hence $I_\mathrm{el}^{(p)}\sim D^{-2(p+1)}$. On the other hand, $d_k\sim D^{-(k+1)}$ and $b_n$ are independent of $D$, hence the sum in Eq.~(\ref{eq:autocorrelation_inel}) is dominated by the term containing $d_0$ and $I_\mathrm{inel}^{(q)}\sim D^{-1}$ for each $q$; therefore inelastic scattering dominates in the strong noise limit. 

\subsection{Two-tone acoustic control of photon scattering}

In this section, we study the RF spectra in the presence of the acoustic modulation composed of two harmonics,
\begin{equation}
\label{eq:two-tone}
\Delta\omega_\mathrm{ac}(t) = A\oma\cos(\oma t) + B\oma\cos(2\oma t+\varphi).
\end{equation}
In the case of such a two-tone modulation one finds 
\begin{align}
b_n^{(2)} = & \sum_{k} J_{n-2k}(A)J_{k}\left(B/2\right)e^{ik\varphi}, \label{eq:bn2}\\
\phi_n^{(2)}(u) = & \sum_k i^{n-k}
J_{n-2k}\left(2A\sin\frac{\oma u}{2}\right)
J_k\left(B\sin\oma u\right) e^{ik\varphi}. \label{eq:fin2}
\end{align}
We will focus on the first acoustic sideband at $\omega=\omega_\mathrm{L}+\oma$ under resonant excitation ($\omega_\mathrm{L}=\overline{\omega}_0$). We will determine the scattering intensity $S(\omega_\mathrm{L}+\oma)$, which corresponds directly to the number of detector counts for the spectral filter set at the first sideband. 

\begin{figure}[tb]
\includegraphics[width=\linewidth]{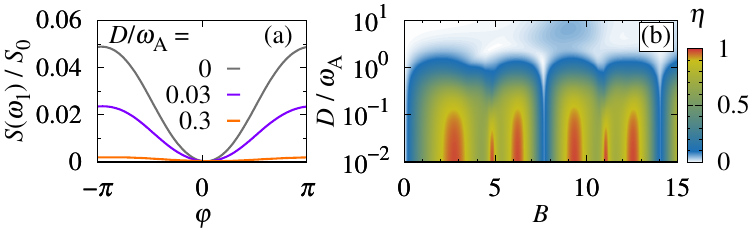}
\caption{\label{fig:interferogram_and_contrast} (a): The scattering intensity at the spectral position of the first acoustic sideband for different noise strengths upon modulation with two commensurate harmonic acoustic waves, as a function of the relative phase between these acoustic harmonics. (b) The contrast of the phase dependence as a function of the amplitude $B$ and the noise strength $D$. Here the QD is excited resonantly, $A=1$ and $B=2.76$ in (a).}
\end{figure}

We calculate the coefficients $b_n$ and $c_n$ numerically and obtain the spectrum from \eqref{eq:RF_spectrum}.
The amplitude of the first sideband as a function of the relative phase $\varphi$ is presented in Fig.~\ref{fig:interferogram_and_contrast}a for different noise strengths $D$. 
Clearly, noise reduces the amplitude of phase-dependent oscillations of the scattering intensity, which is related to the overall intensity reduction discussed above. 

As an intensity-independent figure of merit characterizing the DOC of the scattering intensity we use the normalized contrast $\eta = (S_\mathrm{max}-S_\mathrm{min})/(S_\mathrm{max}+S_\mathrm{min})$, where $S_\mathrm{max}$ and $S_\mathrm{min}$ are the maximum and minimum scattering intensities as a function of the phase $\varphi$. This is shown in Fig.~\ref{fig:interferogram_and_contrast}b as a function of the amplitude $B$ and the noise strength $D$. The contrast approaches unity when the intensity for a certain phase is close to zero. Whether this happens depends on the interplay of various Fourier components in Eq.~(\ref{eq:fin2}). In general, for the intensity to reach zero, at least two of these components must be of comparable order, which occurs at certain values of $B$, corresponding to the red areas in Fig.~\ref{fig:interferogram_and_contrast}(b). The striking apparent periodicity of this picture as a function of $B$ follows from the oscillating character of Bessel functions at large values of their argument. The direct dependence on $J_k(B/2)$ is explicit in \eqref{eq:bn2}, while in \eqref{eq:fin2} it can be shown by appropriately transforming the formula (see Supplement 1). 
The detrimental effect of noise is stronger whenever high relative contrast is due to a very low value of $S_\mathrm{min}$, which makes it more vulnerable to the raising inelastic background.

\begin{figure}[tb]
\includegraphics[width=\linewidth]{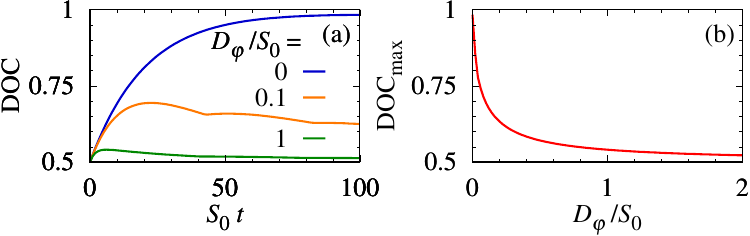}
\caption{\label{fig:fidelity}Impact of the acoustic phase instability on the DOC of the phonon scattering at $B = 2.76$: (a) DOC as a function of the integration time. (b) Maximum achievable DOC.}
\end{figure}

Finally, we discuss the importance of the phase stability of the acoustic modulation for the efficient control of photon scattering. The optical signal from a single quantum emitter has to be integrated over a finite time to obtain meaningful results, and phase instability on this time scale affects the degree of the acoustic phase control of the scattering. As a figure of merit that captures both the accumulation of the physical signal in time and the phase perturbation, we propose DOC defines as follows, which quantifies the amount of phase information encoded in scattered photons: The phase is randomly set to the value that yields the maximum or minimum scattering rate to the first sideband. This setting is to be determined on the basis of the number of photons scattered during a time period $t$ at the frequency of the first sideband. The DOC is equal to the probability of correctly determining the phase setting, with the value of $1/2$ corresponding to the null information (random guessing). The formal details are discussed in Supplement 1. For simplicity, in this discussion we assume that the background noise is absent. 

In a perfectly phase-stable setup, the DOC increases as the signal is integrated in time, asymptotically reaching the value of the contrast $\eta$ (blue line in Fig.~\ref{fig:fidelity}(a)). With phase instability, which we model as a phase diffusion with the diffusion constant $D_\varphi$, the phase information initially grows as photons are collected but then starts to decay since the phase diffusion blurs the initial phase setting, suppressing the difference in the corresponding scattering rates (orange and green lines in Fig.~\ref{fig:fidelity}(a)). As a result, the maximum achievable DOC decays very fast (see Fig.~\ref{fig:fidelity}(b)). To give a rough estimate of the absolute numbers, $\Omega\sim 0.1\gamma$ to ensure weak coupling limit and $\Gamma\sim 0.1\oma$ to select the desired sideband. From Fig.~\ref{fig:fidelity}(b) one can see that the $D_\varphi$ must be 1--2 orders lower than $S_0=\Omega^2/(\gamma^2\Gamma)$ for DOC close to 1. Therefore, $D_\varphi \lesssim 10^{-4} \oma$, that is, the acoustic phase must be stable over times 4--5 orders of magnitude longer than the acoustic period (assuming a perfect detector), which highlights the importance of the extremely high stability demonstrated in Ref.~\cite{Weiss2021}.

\section{Conclusion}
\label{sec:conclusions}

We have developed the theory of light scattering on a single quantum emitter with periodically modulated transition energy in the presence of external noise and phase instability of the modulation. By applying this theory to a semiconductor QD modulated by an acoustic field composed of two harmonics and subject to external white noise, we have shown that the achievable degree of acoustic control of the photon scattering in the spectral domain remains very high for appropriate settings of the modulation amplitudes and for noise amplitudes leading to phase diffusion coefficients well below the acoustic modulation frequency. We have also highlighted the importance of the acoustic phase stability over times four to five orders of magnitude longer than the modulation period. 

Our results set the limits for controlling single-photon scattering by classical acoustic waves. In the future, they may offer a starting point for the analysis of quantum information transfer from mechanical to optical qubits in frequency-bin encoding. Natural extensions of our work would be to include noise processes with finite memory, in particular those with super-Ohmic spectral density that precludes a non-trivial short-time limit. 

\begin{backmatter}
\bmsection{Funding} This work has been supported by the  National Science Centre, Poland (NCN) under Grant No. 2023/50/A/ST3/00511 and by the Alexander von Humboldt Foundation under a Research Group Linkage Grant.

\bmsection{Acknowledgments} The authors thank Daniel Groll, Tilmann Kuhn, and Daniel Wigger for fruitful discussions.

\bmsection{Disclosures} The authors declare no conflicts of interest.

\bmsection{Data availability statement} No data were generated or analyzed in the presented research.

\bmsection{Supplemental document} See Supplement 1 for supporting content.

\end{backmatter}
\smallskip


\newpage

\phantom{aa}

\newpage



\section*{Supplementary material}

# Precision of the acoustic control of single photon scattering with semiconductor quantum dots: supplemental document

This is the supplementary material for the article [Link].

**CONTENTS**

## 1. WHITE NOISE AS A LIMIT OF PHYSICAL NOISE PROCESSES

We present here formal arguments that allow us to understand the white-noise model considered in the main text of the paper as a limiting case of certain noise processes of physical relevance.

Let us first consider the noise originating from a bath of oscillators in the classical limit. The oscillators are labeled by $k$ and couple to our two-level system via coupling constants $g_k$, so that the random shift of the upper level energy is

$$\hbar \Delta \omega_{\text{ns}}(t) = 2\hbar \sum_k g_k \left( X_k \cos \omega_k t + Y_k \sin \omega_k t \right),$$

where the quadrature amplitudes $X_k$ and $Y_k$ are independent Gaussian random variables with zero mean and with the variance given by the equipartition of energy, $\sigma_k^2 = \hbar \omega_k / (2 k_\text{B} T)$, where $k_\text{B}$ is the Boltzmann constant and $T$ is the temperature. The autocorrelation function of the fluctuations is then

$$\langle \Delta \omega_{\text{ns}}(t) \Delta \omega_{\text{ns}}(t') \rangle = \sum_k (2 g_k \sigma_k)^2 \cos \omega_k (t - t') = \frac{k_\text{B} T}{\hbar} \int_0^\infty \frac{J(\omega)}{\omega} \cos \omega (t - t') = R(t - t'),$$

where we introduce the spectral density

$$J(\omega) = 4 \sum_k g_k^2 \delta(\omega_k - \omega).$$

For a broad and regular spectral density, the autocorrelation function $R(\tau)$ is short-lived. If it decays faster than any relevant dynamical time scales of the system (exciton decay and modulation in our case) then it may be approximated by a Dirac delta, giving rise to white noise with

$$\langle \Delta \omega_{\text{ns}}(t) \Delta \omega_{\text{ns}}(t + \tau) \rangle = 2 D \delta(\tau)$$

with

$$D = \frac{1}{2} \int_{-\infty}^\infty d\tau R(\tau) = \pi \frac{k_\text{B} T}{\hbar} \lim_{\omega \to 0} \frac{J(\omega)}{\omega}.$$

This is a well-defined limit for an Ohmic reservoir with $J(\omega) \propto \omega$. While such reservoirs are commonly studied and even "paradigmatic" in the theory of open quantum systems, the lattice

(phonon) reservoir in three dimensions is super-Ohmic and therefore remains outside the scope of the present study, even though it is indeed very fast (THz Debye or cut-off frequencies) compared to the time scales relevant here.

As a second, more specific class of noise processes, we consider a cumulative Stark shift due to a large number $N$ of independent charge traps, quickly switching between their two charge states,

$$\Delta\omega_{\text{ns}}(t) = \sum_{k=1}^{N} f_k X_k,$$

where $X_k$ now represent independent two-state Markovian telegraph processes with zero mean and unit variance, and $f_k$ are coupling constants that may depend, e.g., on the distance between the system and the charge trap. We assume that all telegraph processes have the same switching statistics with $\langle X_k(t) X_k(t+\tau) \rangle = e^{-2\lambda\tau}$. Then

$$\langle \Delta\omega_{\text{ns}}(t) \Delta\omega_{\text{ns}}(t+\tau) \rangle = \sum_k f_k^2 e^{-2\lambda\tau}$$

and the process has a short memory if the switching rate $\lambda$ is large. The limit of infinite number of processes is reached by the scaling $f_k = \tilde{f}_k/\sqrt{N}$ with fixed $\tilde{f}_k$ (like in the central limit theorem), so that the physically relevant variance of the energy fluctuations, $\sigma_\omega^2 = \langle [\Delta\omega_{\text{ns}}(t)]^2 \rangle = \sum_k f_k^2$ remains finite. The process is Gaussian if, for any finite sequence of real numbers $\{s_l\}$,

$$\left\langle e^{i \sum_l s_l \Delta\omega_{\text{ns}}(t_l)} \right\rangle = e^{-\sum_{lj} s_l \sigma_{lj} s_j},$$

where

$$\sigma_{lj} = \left\langle \Delta\omega_{\text{ns}}(t_l) \Delta\omega_{\text{ns}}(t_j) \right\rangle = \sigma_\omega^2 \left\langle X_k(t_l) X_k(t_j) \right\rangle$$

are the covariances of the process. For our process, using the independence of the telegraph contributions and expanding in Taylor series, one finds

$$\ln \left\langle e^{i \sum_l s_l \Delta\omega_{\text{ns}}(t_l)} \right\rangle = \sum_k \ln \left[ 1 - \frac{1}{2N} \tilde{f}_k^2 \sum_{lj} s_l \left\langle X_k(t_l) X_k(t_j) \right\rangle s_j + O\left(\frac{1}{N^{3/2}}\right) \right]$$

$$= -\frac{1}{2} \sum_{lj} s_l \sigma_{lj} s_j + O\left(\frac{1}{N^{1/2}}\right).$$

This shows that the process is Gaussian in the limit of a large number of telegraph noise sources, $N \to \infty$, and can be approximated by white noise in the limit of fast switching.

## 2. AUTOCORRELATION FUNCTION

In this section we summarize the derivation of the correlation function given by Eq. (2) of the main text.

The Hamiltonian of the system (in the rotating frame and using rotating wave approximation) reads

$$H(t) = \hbar \left[ \omega_0(t) - \omega_L \right] |1\rangle\langle 1| - \frac{\hbar \Omega}{2} (\sigma_+ + \sigma_-),$$

where $\Omega$ is the optical Rabi frequency and $\sigma_+ = \sigma_-^\dagger = |1\rangle\langle 0|$. The system undergoes spontaneous emission with the rate $\gamma$, described by the Lindblad superoperator

$$L[\rho] = \gamma \left( \sigma_- \rho \sigma_+ - \frac{1}{2} \{\sigma_+ \sigma_-, \rho\} \right),$$

where $\rho$ is the density matrix and $\{,\}$ denotes the anticommutator. The Master Equation governing the evolution of the system has the form

$$\frac{d\rho(t)}{dt} = -\frac{i}{\hbar} [H(t), \rho(t)] + L[\rho(t)]. \tag{S1}$$

Denote the solution of Eq. (S1) by $\rho(t) = \mathfrak{L}_{t_0,t}[\rho(t_0)]$. The Lax quantum regression theorem then yields the autocorrelation function in the form

$$G(t_1, t_2) = \text{Tr} \left( \sigma_- \mathfrak{L}_{t_1, t_2} [\mathfrak{L}_{t_0, t_1} [\tilde{\rho}(t_0)] \sigma_+] \right). \tag{S2}$$

Here $t_0$ is the initial moment of evolution, while $t_1 - t_0$ is a sufficiently long time for the system to reach its steady state. We find the evolution to the leading order in the Rabi frequency $\Omega$ iteratively, following Refs. [1, 2]. Eq. (S1) yields the equations of motion for the elements $\rho_{01}$, $\rho_{10}$, $\rho_{11}$ of the density matrix in the form $\dot{\rho}_{jl} = a_{jl}\rho_{jl} + i\Omega \sum_{mn} b_{jl,mn}\rho_{mn}$, where $a_{11} = -\gamma$, $a_{01} = a_{10}^* = i\Delta - \gamma/2$, and $b_{11,10} = -b_{11,01} = b_{10,11} = -b_{01,11} = 1/2$. The same holds for an arbitrary matrix, not necessarily a density matrix. Since Eq. (S1) is trace-preserving, one has $\rho_{00} = c_0 - \rho_{11}$, where $c_0$ is a constant determined by the initial values ($c_0 = 1$ for a density matrix). In the absence of the laser field ($\Omega = 0$) the equation of motion can be solved trivially to yield the zeroth-order propagation

$$\rho_{jl}^{(0)}(t) = \left[\mathfrak{L}_{t_0,t}^{(0)}\rho(t_0)\right]_{jl} = e^{\int_{t_0}^t ds\, a_{jl}(s)} \rho_{jl}(t_0).$$

In the subsequent orders $r > 0$ in $\Omega$,

$$\rho_{jl}^{(r)}(t) = \left[\mathfrak{L}_{t_0,t}^{(r)}\rho(t_0)\right]_{jl} = i\Omega \int_{t_0}^t ds\, e^{\int_s^t ds'\, a_{jl}(s')} \sum_{mn} b_{jl,mn} \rho_{mn}^{(r-1)}(s).$$

These equations define the perturbative expansion of the evolution superoperator $\mathfrak{L}_{t_1,t_2}$ in powers of $\Omega$. Substituting this evolution into Eq. (S2) one finds, in the leading order of $\Omega^2$, the autocorrelation function in the form [1]

$$G(t_1, t_2) = \frac{\Omega^2}{4} \int_0^\infty du\, e^{-\frac{\gamma}{2}u} e^{-i\Phi(t_1,t_1-u)} \int_0^\infty du'\, e^{-\frac{\gamma}{2}u'} e^{i\Phi(t_2,t_2-u')}, \tag{S3}$$

where

$$\Phi(t_b, t_a) = \int_{t_a}^{t_b} ds\, [\omega_L - \omega_0(s)] = \overline{\Delta}(t_b - t_a) - \Phi_{ac}(t_b, t_a) - \Phi_{ns}(t_b, t_a). \tag{S4}$$

Here

$$\Phi_{ac}(t_b, t_a) = \int_{t_a}^{t_b} ds\, \Delta\omega_{ac}(s), \quad \Phi_{ns}(t_b, t_a) = \int_{t_a}^{t_b} ds\, \Delta\omega_{ns}(s),$$

we changed the variables according to $s = t + u$ and set $t - t_0 \to \infty$ (steady-state regime). Substituting the decomposition given by the rightmost form of Eq. (S4) to Eq. (S3) one arrives at Eq. (3) from the main text upon using the fact that the noise is stationary, hence

$$\overline{e^{i\Phi(t_d+s,t_c+s) \pm i\Phi(t_b+s,t_a+s)}} = \overline{e^{i\Phi(t_d,t_c) \pm i\Phi(t_b,t_a)}}.$$

## 3. NOISE FUNCTION FOR WHITE NOISE

Here we outline the calculations leading to the noise kernel $\mathcal{D}(u, u'\tau)$ in the form of Eq. (4) and Eq. (5) of the main text.

To evaluate the noise kernel given by Eq. (3) of the main text in the case of white noise (phase diffusion), we need to split the two phases $\Phi_{nc}$ accumulated over the time intervals $(-u, 0)$ and $(\tau - u', \tau)$ into non-overlapping time intervals that will allow factorization based on the independence of the corresponding increments of the diffusion process. This is done in different ways in each of the areas (a)–(c) in the $(u, u')$ plane shown in the upper part of Fig. S1. We denote the characteristic functions of these areas by $\Theta_i(u, u'; \tau)$, $i = a, b, c$. In each of the areas, the phases partly cancel in a particular way, which we represent graphically in the lower part of Fig. S1. Here the positive phase is represented by a right-heading arrow above the time axis and the negative phase is shown as a left-heading arrow below, so that each arrow represents an integral of $\Delta\omega_{ns}$ over the interval from its tail to head. The original phases are shown on the left, while the final form after partial cancellation is on the right. For instance, in area (a), the noise kernel has the form

$$\mathcal{D}_a(u, u', \tau) = \overline{e^{-i\Phi_{ns}(-u, \tau-u')}} \, \overline{e^{-i\Phi_{ns}(\tau,0)}}.$$

The phase $\Phi_{ns}(t, t')$ undergoes normal diffusion from time $t'$ to $t$ induced by the white noise $\Delta\omega_{ns}$ and is therefore normally distributed with zero mean and variance $2D(t - t')$. Therefore

$$\overline{e^{i\Phi_{ns}(t,t')}} = e^{-D(t-t')},$$

which leads to

$$\mathcal{D}(u, u', \tau) = \mathcal{D}_a(u, u', \tau)\Theta_a(u, u', \tau) + \mathcal{D}_b(u, u', \tau)\Theta_b(u, u', \tau) + \mathcal{D}_c(u, u', \tau)\Theta_c(u, u', \tau)$$

3

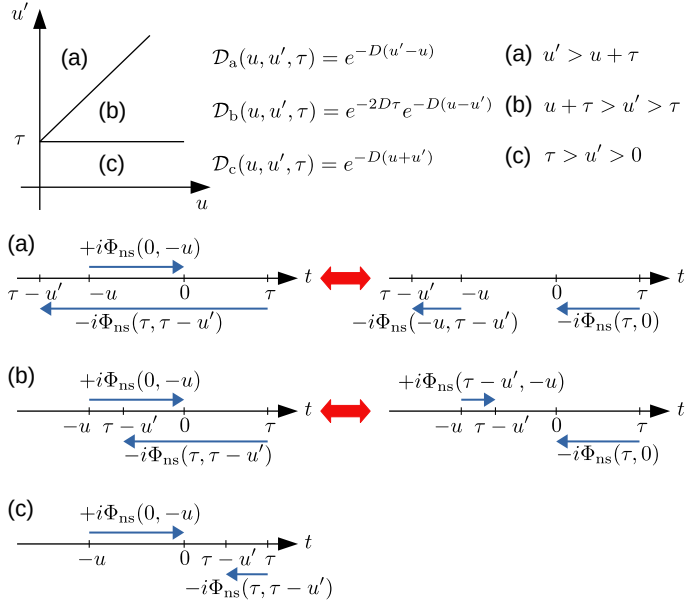

**Fig. S1.** Upper: division of the $(u, u')$ plane into areas for the calculation of the noise kernel. Lower: Graphical representation of the reduction of the phase evolution intervals into non-overlapping parts.

with

$$\mathcal{D}_a(u, u', \tau) = e^{-D(u'-u)}, \quad \mathcal{D}_b(u, u', \tau) = e^{-2D\tau}e^{-D(u-u')}, \quad \mathcal{D}_c(u, u', \tau) = e^{-D(u+u')}.$$

From this, we can extract a part that is undamped in $\tau$,

$$\mathcal{D}_{el}(u, u', \tau) = \mathcal{D}_c(u, u', \tau)\left[\Theta_a(u, u', \tau) + \Theta_b(u, u', \tau) + \Theta_c(u, u', \tau)\right] = \mathcal{D}_c(u, u', \tau)\theta(u)\theta(u'),$$

which is the contribution given by Eq. (4) of the main text. The remaining part

$$\mathcal{D}_{inel}(u, u', \tau) = \left[\mathcal{D}_b(u, u', \tau) - \mathcal{D}_c(u, u', \tau)\right]\Theta_b(u, u', \tau) + \left[\mathcal{D}_a(u, u', \tau) - \mathcal{D}_c(u, u', \tau)\right]\Theta_a(u, u', \tau),$$

can be written in the form of Eq. (5) of the main text.

## 4. AUTOCORRELATION FUNCTION: INELASTIC PART

We present here some details of the derivation of Eq. (9) from the main text of the paper.

Since the phase $\Phi_{ac}(t, t')$ is defined as a definite integral, one can rearrange the phase factors in Eq. (2) of the main text,

$$\Phi_{ac}(t_1, t_1 - u) - \Phi_{ac}(t_2, t_2 - u') = -\Phi_{ac}(t_2, t_1) + \Phi_{ac}(t_2 - u', t_1 - u).$$

Next, performing the change of variables $u = (x+y)/2$, $u' = t_2 - t_1 + (x-y)/2$, $-\infty < x < \infty$, $|y| < x < \infty$, and using the expansion from Eq. (6) in the main text, one can write Eq. (2) in the form

$$G_{inel}(t_1, t_2) = \frac{\Omega^2}{8} e^{-\left(\frac{\gamma}{2} + D - i\bar{\Delta}\right)(t_2 - t_1)} e^{-i\Phi_{ac}(t_2, t_1)} \sum_k e^{ik\omega_A t_1} \int_{-\infty}^{\infty} dy \phi_k(y) e^{-i\bar{\Delta}y}$$
$$\times \int_{|y|}^{\infty} dx e^{-\frac{\gamma + ik\omega_A}{2}x} \left[e^{-D|y|} - e^{-Dx}\right]. \quad (S5)$$

Performing trivial integration over $x$, then employing Eq. (8) for the integration over $y$ and using Eq. (10) one finds the result in the form of Eq. (9) in the main text.

4

## 5. EXPLICIT FORM OF THE RESONANCE FLUORESCENCE SPECTRUM

The spectrum follows in a straightforward way from the exponential $\tau$-dependence in Eq. (7). We present the formulas here for completeness.

Substituting $t_1 = t$, $t_2 = t + \tau$ to Eq. (7), integrating over $t$ according to Eq. (13) and then substituting to Eq. (12) immediately yields the elastic part of the spectrum in the form

$$S_{\rm el}(\omega) = F_0 \operatorname{Re} \sum_n \frac{\Gamma |c_n|^2}{\Gamma - i(\omega - \omega_L - n\omega_{\rm A})}. \tag{S6}$$

In the same way, using Eq. (9) we get the inelastic contribution to the spectrum,

$$S_{\rm inel}(\omega) = \frac{F_0}{2} \operatorname{Re} \sum_{n,k} \frac{\Gamma b_n^* b_{n-k} d_k}{\Gamma + \frac{\gamma}{2} + D - i(\omega - \overline{\omega}_0 - n\omega_{\rm A})}. \tag{S7}$$

## 6. TWO-TONE AND SINGLE-TONE MODULATION; ASYMPTOTICS IN THE STRONG-NOISE LIMIT

In this section we derive Eqs. (14), (18) and (19) from the main text. Then we explain how the power-law exponents discussed in the final part of Sec. 3A can be derived.

For a two-tone modulation as in Eq. (16) of the main text, we write the phase in the form

$$i\Phi_{\rm ac}(t - u', t - u) = 2iA \sin\left[\frac{\omega_{\rm A}(u - u')}{2}\right] \cos\left[\omega_{\rm A} t - \frac{\omega_{\rm A}(u + u')}{2}\right]$$
$$+ iB \sin\left[\omega_{\rm A}(u - u')\right] \cos\left[2\omega_{\rm A} t - \omega_{\rm A}(u + u') + \varphi\right].$$

Using Jacobi-Anger expansion with respect to the cos terms, while treating the sin terms as coefficients, leads to the expansion in Eq. (6) with the functions $\phi_n$ given by Eq. (19). The expansion in Eq. (10) with the coefficients given by Eq. (18) is obtained by writing the definite integral $\Phi_{\rm ac}(t_2, t_1)$ as a difference of originals at the final and initial points and directly applying the Jacobi-Anger expansion. The results for a single-tone harmonic modulation are retrieved by setting $B = 0$.

Functions $\phi_n(u)$ can be written in an alternative form. Let us focus on the factor dependent on $B$ in Eq. (19) (the argument is the same for the other factor). The Bessel function can be written in terms of the defining integral

$$J_k\left(B \sin \omega_{\rm A} u\right) = \frac{1}{2\pi} \int_{-\infty}^{\infty} dt\, e^{ikt} e^{-iB \sin \omega_{\rm A} u \sin t}.$$

We write $B \sin \omega_{\rm A} u \sin t = B[\cos(\omega_{\rm A} u - t) + \cos(\omega_{\rm A} u + t)]/2$, apply the Jacobi-Anger expansion twice, and perform the resulting trivial integration over $t$. The result is

$$J_k\left(B \sin \omega_{\rm A} u\right) = i^{-k} \sum_m J_{m+k}(B/2) J_m(B/2) e^{i(2m+k)\omega_{\rm A} u}.$$

For moderate modulation amplitudes, only a few Bessel functions contribute. All of them oscillate with the same period when $B$ is sufficiently large. Hence, $J_k(B \sin \omega_{\rm A} u)$ also oscillates in $B$ and so does $\phi_n(u)$. Upon substituting this expression, along with the analogous expression for $J_{n-2k}(2A \sin \omega_{\rm A} u/2)$, to Eq. (19) and then to Eq. (8), the integration can be performed analytically and one obtains a summation formula that can be evaluated numerically much faster than the original integral.

To obtain the asymptotic behavior of the scattering intensities for the single-tone modulation in the limit of strong noise, as shown in Fig. 2 of the main text and discussed in Sec. 3A, we need to extract the asymptotics of the coefficients $c_n$ and $d_n$ as $D \to \infty$ (note that $b_n$ does not depend on $D$). Substituting $\phi_n^{(1)}(u)$ from Eq. (14) to Eq. (8) and changing the variable to $x = uD$ we obtain

$$c_n^{(1)} = \frac{\gamma i^n}{2D} \int_0^\infty dx\, e^{-\frac{\gamma + 2D}{2D} x} e^{-i\frac{\overline{\Delta} + n\omega_{\rm A}/2}{D} x} J_n\left[2A \sin\left(\frac{\omega_{\rm A}}{2D} x\right)\right] \approx \frac{\gamma i^n}{2D} \int_0^\infty dx\, e^{-x} J_n\left(\frac{A\omega_{\rm A}}{D} x\right),$$

where we assumed $D \gg \gamma, \omega_{\rm A}, \overline{\Delta}$ in the final expression. Substituting the leading term for the Jacobi function, $J_n(z) \approx z^n/(2^n n!)$, $z \ll 1$, we get

$$c_n^{(1)} \approx \frac{\gamma}{2D} \left(\frac{iA\omega_{\rm A}}{2D}\right)^n \sim D^{-(n+1)}.$$

Then, from Eq. (11), also $d_n \sim D^{-(n+1)}$.

## 7. ACOUSTIC PHASE INFORMATION IN THE SCATTERING SPECTRUM

We want to develop a quantitative measure of the degree of control (DOC) of light scattering by the relative phase between two harmonic components of acoustic modulation. Scattering is efficiently controlled by this acoustic phase if the scattering intensity depends on the phase setting, which means that the phase information is encoded in the scattering spectrum. We therefore propose to quantify the DOC in terms of a single-shot attempt to determine the acoustic phase from the accumulated scattering data. We choose the simplest scheme in which the detector is spectrally tuned to the first spectral sideband and the phase is randomly set with equal probabilities to one of the values $\varphi_1$ or $\varphi_2$, corresponding, respectively, to the maximum and minimum of the scattering intensity at this spectral position (see Fig. 3 of the main text). Roughly speaking, the scattering is controlled by the acoustic phase if the detection counts accumulated over a certain period of time allow one to determine whether the scattering intensity is "high" or "low" and thus correctly infer the initial setting of the phase with a high probability. More precisely, the signal is integrated over a period of time $T$ and one opts for $\varphi_1$ if the number of counts $N$ is above a threshold value $\nu$. The probability of wrong inference is then given by

$$P_\text{e} = p\left(N \leq \nu | \varphi = \varphi_1\right) p\left(\varphi = \varphi_1\right) + p\left(N > \nu | \varphi = \varphi_2\right) p\left(\varphi = \varphi_2\right) \quad \text{(S8)}$$
$$= \frac{1}{2}\left[p\left(N < \nu | \varphi = \varphi_1\right) + p\left(N > \nu | \varphi = \varphi_2\right)\right].$$

The degree of phase control is defined as the probability of correctly determining the phase for an optimal choice of $\nu$.

$$\text{DOC} = \sup_\nu \left(1 - P_\text{e}\right).$$

For the sake of our discussion it is convenient to treat the measurement time as a parameter over which the procedure needs to be optimized in the final step.

With absolutely stable acoustic phase, $p(N = n | \varphi = \varphi_i)$, $i = 1, 2$, are given by Poisson distribution with the distribution parameter (mean count number) $\lambda_i(T) = w_i T$, where $w_i$ are the scattering rates for the two phase settings, proportional to the scattering intensities $I_1 = I_\text{max}$ and $I_2 = I_\text{min}$. With phase instability, the phase diffuses so that at a time $t$ it is distributed according to the probability density $f_i(\varphi, t)$, $f_i(\varphi, t = 0) = \delta(\varphi - \varphi_i)$. Here we will assume normal diffusion with the diffusion constant $D_\varphi$,

$$f_i(\varphi, t) = \frac{1}{\sqrt{2\pi D_\varphi t}} e^{-\frac{(\varphi - \varphi_i)^2}{2 D_\varphi t}}.$$

As a result, while the count distribution remains Poissonian, the scattering rate becomes time-dependent,

$$w_i(t) = \int_{-\infty}^{\infty} d\varphi\, w(\varphi) f_i(\varphi, t), \quad \lambda_i(T) = \int_0^T d\tau\, w_i(\tau),$$

where $w(\varphi)$ is the scattering rate for the phase set to $\varphi$.

From Eq. (S8) one has

$$P_\text{e} = \frac{1}{2} e^{-\lambda_1} \sum_{n=0}^{\lfloor \nu \rfloor} \frac{\lambda_1^n}{n!} + \frac{1}{2} e^{-\lambda_2} \sum_{n=\lfloor \nu \rfloor + 1}^{\infty} \frac{\lambda_2^n}{n!},$$

where $\lfloor \nu \rfloor$ denotes the largest integer not greater than $\nu$. This has a minimum with respect to $\nu$ when the two distributions are equal,

$$e^{-\lambda_1} \frac{\lambda_1^n}{n!} = e^{-\lambda_2} \frac{\lambda_2^n}{n!} \text{ for } n = \nu.$$

which yields

$$\nu = 2\lambda\eta \left(\ln \frac{1+\eta}{1-\eta}\right)^{-1}, \quad \lambda = \frac{\lambda_1 + \lambda_2}{2}, \quad \eta = \frac{\lambda_1 - \lambda_2}{\lambda_1 + \lambda_2}.$$

The resulting DOC is

$$\text{DOC} = \frac{1}{2} + \frac{1}{2} e^{-\lambda} \left(1 - \eta^2\right)^{\nu/2} \sum_{n=\lfloor \nu \rfloor + 1}^{\infty} \frac{\lambda^n}{n!} \left[(1+\eta)^{n-\nu} - (1-\eta)^{n-\nu}\right],$$

which depends on the measurement time via the quantities $\lambda$ and $\eta$.

6



\end{document}